\documentclass[12pt]{article}

\catcode`@=12
\begin{document}
\thispagestyle{empty}
\begin{flushright} 
UCRHEP-T350\\ 
December 2002\
\end{flushright}
\vspace{0.5in}
\begin{center}
{\LARGE	\bf Gauge Model of Quark-Lepton Nonuniversality\\}
\vspace{1.5in}
{\bf Xiao-Yuan Li$^a$ and Ernest Ma$^b$\\}
\vspace{0.2in}
{$^a$ \sl Institute of Theoretical Physics, Chinese Academy of Sciences, 
Beijing, China\\}
\vspace{0.1in}
{$^b$ \sl Physics Department, University of California, Riverside, 
California 92521, USA\\}
\vspace{1.5in}
\end{center}
\begin{abstract}\
We propose a gauge model where quark-lepton universality is an accidental 
symmetry which is only approximate, in analogy to the well-accepted notion 
that strong isospin is accidental and approximate.  This is a natural 
framework for explaining possible small deviations of quark-lepton 
universality which is applicable to the recently reported apparent 
nonunitarity of the quark mixing matrix.  As a result, small departures 
from quark-lepton universality are expected in $Z$ decays as well 
as in the recent neutrino data of the NuTeV collaboration and in future 
low-energy experiments.  New physics is predicted at the TeV scale.
\end{abstract}
\newpage
\baselineskip 24pt

In the standard model of particle interactions, left-handed (right-handed) 
quarks and leptons are doublets (singlets) under the same $SU(2)_L \times 
U(1)_Y$ gauge group.  This has three important implications: (1) the observed 
weak-interaction strength of quarks is equal to that of leptons, i.e. $G_F^q 
= G_F^l$, (2) the observed weak-interaction strength of each generation of 
quarks and leptons is equal to one another, i.e. $G_F^e = G_F^\mu = G_F^\tau$, 
and (3) the charged-current strength is equal to the neutral-current strength, 
i.e. $G_F^{CC} = G_F^{NC}$, if the $SU(2)_L \times U(1)_Y$ gauge symmetry is 
spontaneously broken only by scalar doublets.

Experimentally, all seem to be in very good agreement with data, but as more 
precision data become available, it is theoretically desirable to have a 
natural framework for describing any possible deviations.  We should use 
what we have learned regarding the validity of strong isospin, which is now 
understood as an accidental approximate symmetry because of the mass-scale 
hierarchy $m_u, m_d << \Lambda_{QCD}$ and not because $m_u=m_d$ (or more 
precisely $|m_u - m_d| << m_u + m_d$) as previously thought.  We are thus 
motivated to propose that quarks and leptons couple to \underline {different} 
$SU(2)$'s and $U(1)$'s with \underline {different} coupling strengths, but 
their effective low-energy weak-interaction strengths will turn out to be 
\underline {independent} of the couplings, and are nearly equal because of a 
certain mass-scale hierarchy of scalar vacuum expectation values, 
i.e. $G_F^q \simeq G_F^l$.  This remarkable result was first 
obtained over 20 years ago \cite{lima81} and applied to generation 
nonuniversality \cite{lima81,malitu88,mang88, lima92,lima93}.  Just as this 
previous proposal required that the $\tau$ lifetime be longer than the 
standard-model prediction (which is no longer supported by present data), 
our new proposal of quark-lepton nonuniversality requires that the neutron 
lifetime be longer, which is exactly what has now been reported \cite{neutron}.

This recent measurement of the neutron $\beta-$decay asymmetry has determined
that
\begin{equation}
|V_{ud}| = 0.9713(13),
\end{equation}
which, together with \cite{pdg} $|V_{us}| = 0.2196(23)$ and $|V_{ub}| = 
0.0036(9)$, implies the apparent nonunitarity of the quark mixing matrix, i.e.
\begin{equation}
|V_{ud}|^2 + |V_{us}|^2 + |V_{ub}|^2 = 0.9917(28).
\end{equation}
However, if the effective $G_F^{CC}$ measured in lepton-quark interactions 
is smaller than that measured in lepton-lepton interactions, i.e. $\mu$ 
decay, then the above is \underline {expected} to be less than one.

Consider the gauge group $SU(3)_c \times SU(2)_q \times SU(2)_l \times U(1)_q 
\times U(1)_l$.  This differs from our previous proposals by the additional 
extension of $U(1)_Y$ to $U(1)_q \times U(1)_l$.  
Quarks and leptons are assumed to transform as follows:
\begin{eqnarray}
&& (u,d)_L \sim (3,2,1,1/6,0), ~~ u_R \sim (3,1,1,2/3,0), ~~ d_R \sim (3,1,1,
-1/3,0); \\ && (\nu,e)_L \sim (1,1,2,0,-1/2), ~~ e_R \sim (1,1,1,0,-1).
\end{eqnarray}
The scalar sector consists of two doublets
\begin{equation}
(\phi_1^+,\phi_1^0) \sim (1,2,1,1/2,0), ~~ (\phi_2^+,\phi_2^0) \sim (1,1,2,0,
1/2), 
\end{equation}
one singlet
\begin{equation}
\chi^0 \sim (1,1,1,1/2,-1/2),
\end{equation}
and one self-dual bidoublet
\begin{equation}
\eta = {1 \over \sqrt 2} \pmatrix {\eta^0 & -\eta^+ \cr \eta^- & \bar \eta^0} 
\sim (1,2,2,0,0),
\end{equation}
such that $\eta = \tau_2 \eta^* \tau_2$. Each column is a doublet under 
$SU(2)_q$ and each row is a doublet under $SU(2)_l$.
 
Let $\langle \phi^0_{1,2} \rangle \equiv v_{1,2}$, $\langle \chi^0 \rangle 
\equiv w$, and $\langle \eta^0 \rangle \equiv u$, then the effective 
charged-current four-fermion weak coupling strengths at low energy are given 
by \cite{lima93}
\begin{eqnarray}
\left( {4 G_F \over \sqrt 2} \right)^{CC}_{ll} &=& {u^2 + v_1^2 \over (v_1^2 + 
v_2^2) u^2 + v_1^2 v_2^2} = \left( {4 G_F \over \sqrt 2} \right)_\mu, \\ 
\left( {4 G_F \over \sqrt 2} \right)^{CC}_{lq} &=& {u^2 \over (v_1^2 + v_2^2) 
u^2 + v_1^2 v_2^2} = \left( {4 G_F \over \sqrt 2} \right)_\beta.
\end{eqnarray}
Note that these expressions are independent of the $SU(2)_q$ and $SU(2)_l$ 
couplings and quark-lepton universality is obtained in the limit $v_{1,2}^2 
<< u^2$.  The effective $G_F$ measured in $d \to u + e + \bar \nu_e$  
\underline {must} now be smaller than that in $\mu \to \nu_\mu + e + \bar 
\nu_e$ by the factor $\xi^{-1}$, where \cite{lima81}
\begin{equation}
\xi \equiv 1 + {v_1^2 \over u^2}.
\end{equation}
Thus the apparent unitarity violation of the quark mixing matrix [Eq.~(2)] 
can be explained with
\begin{equation}
{v_1^2 \over u^2} = 0.0042(14).
\end{equation}
This potential effect was already pointed out over 10 years ago \cite{mara90}, 
in response to a proposed model \cite{gjs} where $\xi=1$ exactly, in 
which case there is no such effect.

Consider now the effective neutral-current four-fermion weak coupling 
strengths. They are given by \cite{lima93}
\begin{eqnarray}
\left( {4 G_F \over \sqrt 2} \right)^{NC}_{lq} &=& {u^2 w^2 \over (v_1^2 + 
v_2^2) u^2 w^2 + v_1^2 v_2^2 (u^2 + w^2)}, \\ 
\left( {4 G_F \over \sqrt 2} \right)^{NC}_{ll} &=& {u^2 w^2 + v_1^2 (u^2 + w^2)
\over (v_1^2 + v_2^2) u^2 w^2 + v_1^2 v_2^2 (u^2 + w^2)}.
\end{eqnarray}
This shows that the effective neutral-current $G_F$ given in Eq.~(12) 
measured in low-energy neutrino-quark scattering should be smaller than the 
corresponding charged-current $G_F$ given in Eq.~(9) by the factor
\begin{equation}
{(G_F)^{NC}_{lq} \over (G_F)^{CC}_{lq}} = \left( 1 + {v_1^2 v_2^2 u^2 \over 
w^2 [(v_1^2 + v_2^2) u^2 + v_1^2 v_2^2]} \right)^{-1} \simeq 1 - \left( 
{v_1^2 v_2^2 \over v_1^2 + v_2^2} \right) {1 \over w^2}.
\end{equation}
Similarly, the effective $\sin^2 \theta_W$ will also be shifted.  We thus 
expect small deviations from the Standard Model in precision low-energy 
neutrino-quark scattering experiments such as NuTeV \cite{nutev}.  On the 
other hand, the size of these deviations is constrained by the structure 
of the gauge model and is not enough to explain the recent NuTeV result. 
We will come back to this after we consider the precision data at the $Z$ 
resonance.

Let $g_{1,2,3,4}$ be the gauge couplings of $SU(2)_q$, $SU(2)_l$, $U(1)_q$, 
$U(1)_l$ respectively and define $g_{ij}^{-2} \equiv g_i^{-2} + g_j^{-2}$. 
The electromagnetic coupling $e$ is then given by
\begin{equation}
{1 \over e^2} = {1 \over g_1^2} + {1 \over g_2^2} + {1 \over g_3^2} + 
{1 \over g_4^2},
\end{equation}
and the photon in the basis $(W^0_q, W^0_l, B_q, B_l)$ is
\begin{equation}
A = e(g_1^{-1}, g_2^{-1}, g_3^{-1}, g_4^{-1}).
\end{equation}
We now consider the following 3 orthonormal states:
\begin{eqnarray}
Z_1 &=& e(g_{12} g_{34}^{-1} g_1^{-1}, g_{12} g_{34}^{-1} g_2^{-1}, 
-g_{34} g_{12}^{-1} g_3^{-1}, -g_{34} g_{12}^{-1} g_4^{-1}), \\ 
Z_2 &=& g_{12} (g_2^{-1}, -g_1^{-1}, 0, 0), \\ 
Z_3 &=& g_{34} (0, 0, g_4^{-1}, -g_3^{-1}).
\end{eqnarray}
The resulting $3 \times 3$ mass-squared matrix is then given by
\begin{equation}
{1 \over 2} \pmatrix {(g_{12}^2 g_{34}^2 /e^2) (v_1^2 + v_2^2) 
& (g_{12}^2 g_{34} /e g_1 g_2) (g_1^2 v_1^2 - g_2^2 v_2^2) & 
(g_{12} g_{34}^2 /e g_3 g_4) (g_4^2 v_2^2 - g_3^2 v_1^2) \cr 
(g_{12}^2 g_{34} /e g_1 g_2) (g_1^2 v_1^2 - g_2^2 v_2^2) & 
(g_1^2 g_2^2 /g_{12}^2) u^2 + {\cal O}(v^2) & {\cal O}(v^2) \cr 
(g_{12} g_{34}^2 /e g_3 g_4) (g_4^2 v_2^2 - g_3^2 v_1^2) & 
{\cal O}(v^2) & (g_3^2 g_4^2 /g_{34}^2) w^2 + {\cal O}(v^2)}.
\end{equation}
Whereas $Z_1$ couples universally to quarks and leptons, $Z_2$ and $Z_3$ 
will distinguish between them.  The observed $Z$ boson is mostly $Z_1$ 
with small mixtures of $Z_2$ and $Z_3$ of order $v^2/u^2$ and $v^2/w^2$ 
respectively.

Define $r \equiv v_2^2/v_1^2$, $y \equiv g_2^2/(g_1^2+g_2^2)$, $x \equiv 
g_4^2/(g_3^2+g_4^2)$, and consider the leptonic decay width
\begin{equation}
\Gamma_l = {G_F m_Z^3 \over 24 \sqrt 2 \pi} \left( 1 + {3 \alpha \over 4 \pi} 
\right) \rho_l [ 1 + (1-4\sin^2 \theta_l)^2 ]
\end{equation}
with the analogous expression for quarks, then a straightforward analysis 
\cite{lima93} yields
\begin{eqnarray}
\Delta \rho_l &=& - y^2 (1+r){v_1^2 \over u^2} +  {[1 - x^2 
(1+r)^2] \over 1+r}{v_1^2 \over w^2}, \\ 
\Delta \rho_q &=& \Delta \rho_l - 2 [1-y(1+r)] {v_1^2 \over u^2} - 
2 [1-x(1+r)] {v_1^2 \over w^2}, \\ 
\Delta \sin^2 \theta_l &=&  {s^2 y [1 - s^2 y(1+r)] \over c^2-s^2} {v_1^2 
\over u^2} +  {c^2 [1-x(1+r)] \over (c^2-s^2) (1+r)} 
[-s^2+c^2x(1+r)]{v_1^2 \over w^2}, \\ 
\Delta \sin^2 \theta_q &=& \Delta \sin^2 \theta_l + s^2 [1-y(1+r)] {v_1^2 
\over u^2} - c^2 [1-x(1+r)] {v_1^2 \over w^2},
\end{eqnarray}
where $c^2 = 1 - s^2$ and $s^2$ may be taken to be any one of the several 
$\sin^2 \theta_W$ values measured in various ways (because their differences 
would be of higher order in the correction).  Note that $\Delta \rho_q - 
\Delta \rho_l$ and $\Delta \sin^2 \theta_q - \Delta \sin^2 \theta_l$ shift 
in the same direction from $Z_1-Z_3$ mixing but in opposite directions from 
$Z_1-Z_2$ mixing.  This allows for the possibility that $\Delta \rho_q$ and 
$\Delta \rho_l$ are equal, but $\Delta \sin^2 \theta_q$ and $\Delta \sin^2 
\theta_l$ are not, or vice versa.  Using the above shifts 
and the method of Ref.~\cite{bgklm}, we may then write down the deviations 
expected from the standard model for all the precision measurements at the 
$Z$ resonance \cite{pdg}.

In low-energy neutrino-quark scattering, the effective neutral-current 
interaction is given by
\begin{equation}
{\cal H}_{int} = \left( {4 G_F \over \sqrt 2} \right)^{NC}_{lq} {1 \over 2} 
\bar \nu \gamma^\mu \left( {1-\gamma_5 \over 2} \right) \nu \left[ 
j^{(3)}_{qL} - (\sin^2 \theta_W)_{lq} j^{em}_q \right]_\mu,
\end{equation}
where
\begin{equation}
(\sin^2 \theta_W)_{lq} = {e^2 \over g_{12}^2} + {e^2 \over g_2^2} {v_1^2 \over 
u^2} - {e^2 \over g_4^2} {v_1^2 \over w^2}.
\end{equation}
Using Eq.~(20), we obtain
\begin{equation}
{e^2 \over g_{12}^2} = s_0^2 \left[ 1 + {c^2 \over (c^2-s^2)} {1 \over 1+r} 
\left[ \left( 1-[1-y(1+r)]^2 \right) {v_1^2 \over u^2} - [1-x(1+r)]^2 
{v_1^2 \over w^2} \right] \right],
\end{equation}
where $s_0^2$ is defined as usual by
\begin{equation}
s_0^2 (1-s_0^2) = {\pi \alpha (M_Z) \over \sqrt 2 G_F M_Z^2}.
\end{equation}
Thus the shift of $(\sin^2 \theta_W)_{lq}$ from the standard-model prediction 
using precision data at the $Z$ resonance is given by
\begin{eqnarray}
(\Delta \sin^2 \theta_W)_{lq} &=& \left[ {s^2 c^2 \over (c^2-s^2)} [2y - 
y^2(1+r)] + s^2 (1-y) \right] {v_1^2 \over u^2} \nonumber \\ 
&-& \left[ {s^2 c^2 \over (c^2-s^2)} {[1-x(1+r)]^2 \over 1+r} + c^2 (1-x) 
\right] {v_1^2 \over w^2}.
\end{eqnarray}

Using Eq.~(14), we then have
\begin{eqnarray}
\Delta (g_L^{eff})^2 &=& - \left( {2r \over 1+r} \right) {v_1^2 \over w^2} 
(g_L^{eff})^2_{SM} - \left( 1 - {10s^2 \over 9} \right) (\Delta \sin^2 
\theta_W)_{lq}, \\
\Delta (g_R^{eff})^2 &=& - \left( {2r \over 1+r} \right) {v_1^2 \over w^2} 
(g_R^{eff})^2_{SM} + {10s^2 \over 9} (\Delta \sin^2 \theta_W)_{lq},
\end{eqnarray}
as deviations from the standard model appropriate for the NuTeV measurements.

Parity nonconservation in atomic transitions is governed by the same 
effective low-energy neutral-current interaction.  The shift in the weak 
charge is given here by
\begin{equation}
\Delta Q_W = \left( {\Delta G_F \over G_F} \right)^{NC}_{lq} (Q_W)_{SM} 
- 4 Z (\Delta \sin^2 \theta_W)_{lq},
\end{equation}
where
\begin{equation}
\left( {\Delta G_F \over G_F} \right)^{NC}_{lq} = -{v_1^2 \over u^2} - 
\left( {r \over 1+r} \right) {v_1^2 \over w^2},
\end{equation}
which is of the same order of magnitude as the experimental accuracy.

For neutral-current leptonic process such as $\nu_\mu e \to \nu_\mu e$, 
we use Eqs.~(8) and (13) to obtain
\begin{equation}
\left ( {\Delta G_F \over G_F} \right)^{NC}_{ll} = \left( {1 \over 1+r} 
\right) {v_1^2 \over w^2},
\end{equation}
and the analogs of Eqs.~(27) and (30) are now
\begin{equation}
(\sin^2 \theta_W)_{ll} = {e^2 \over g_{12}^2} - {e^2 \over g_1^2} {v_1^2 \over 
u^2} + {e^2 \over g_3^2} {v_1^2 \over w^2},
\end{equation}
and
\begin{eqnarray}
(\Delta \sin^2 \theta_W)_{ll} &=& \left[ {s^2 c^2 \over (c^2-s^2)} [2y - 
y^2(1+r)] - s^2 y \right] {v_1^2 \over u^2} \nonumber \\ 
&-& \left[ {s^2 c^2 \over (c^2-s^2)} {[1-x(1+r)]^2 \over 1+r} - c^2 x 
\right] {v_1^2 \over w^2}.
\end{eqnarray}

We now perform a global fit to all available experimental observables (22 in 
number) and obtain the following best-fit values:
\begin{equation}
{v_1^2 \over u^2} = 0.00489, ~~~ {v_1^2 \over w^2} = 0.00238, ~~~ r = 10.2, 
~~~ y = 0.0955, ~~~ x = 0.135.
\end{equation}
Our results are summarized in Table 1.  Details will be given in a forthcoming 
comprehensive paper.

\begin{table}[htb]
\caption{Fit Values of 22 Observables}
\begin{center}
\begin{tabular}{|c|c|c|c|c|c|}
\hline 
Observable & Measurement & Standard Model & Pull & This Model & Pull \\ 
\hline
$\Gamma_l$ [MeV] & $83.985 \pm 0.086$ & 84.015 & $-0.3$ & 83.950 & $+0.4$ \\ 
$\Gamma_{inv}$ [MeV] & $499.0 \pm 1.5$ & 501.6 & $-1.7$ & 501.2 & $-1.5$ \\ 
$\Gamma_{had}$ [GeV] & $1.7444 \pm 0.0020$ & 1.7425 & $+1.0$ & 1.7444 & $-0.0$ 
\\ 
$A^{0,l}_{fb}$ & $0.01714 \pm 0.00095$ & 0.01649 & $+0.7$ & 0.01648 & $+0.7$ 
\\ 
$A_l(P_\tau)$ & $0.1465 \pm 0.0032$ & 0.1483 &$-0.6$ & 0.1482 & $-0.5$ \\ 
$R_b$ & $0.21644 \pm 0.00065$ & 0.21578 & $+1.0$ & 0.21582 & $+1.0$ \\ 
$R_c$ & $0.1718 \pm 0.0031$ & 0.1723 & $-0.2$ & 0.1722 & $-0.1$ \\ 
$A_{fb}^{0,b}$ & $0.0995 \pm 0.0017$ & 0.1040 & $-2.6$ & 0.1039 & $-2.6$ \\ 
$A_{fb}^{0,c}$ & $0.0713 \pm 0.0036$ & 0.0743 & $-0.8$ & 0.0740 & $-0.8$ \\ 
$A_b$ & $0.922 \pm 0.020$ & 0.935 & $-0.7$ & 0.934 & $-0.6$ \\ 
$A_c$ & $0.670 \pm 0.026$ & 0.668 & $+0.1$ & 0.665 & $+0.2$ \\ 
$A_l$(SLD) & $0.1513 \pm 0.0021$ & 0.1483 & $+1.4$ & 0.1482 & $+1.5$ \\ 
$\sin^2 \theta^{lept}_{eff}(Q_{fb})$ & $0.2324 \pm 0.0012$ & 0.2314 & $+0.8$ 
& 0.2322 & $+0.2$ \\
$m_W$ [GeV] & $80.449 \pm 0.034$ & 80.394 & $+1.6$ & 80.390 & $+1.7$ \\ 
$\Gamma_W$ [GeV] & $2.139 \pm 0.069$ & 2.093 & $+0.7$ & 2.093 & $+0.7$ \\ 
$g_V^{\nu e}$ & $-0.040 \pm 0.015$ & $-0.040$ & $-0.0$ & $-0.039$ & $-0.1$ \\ 
$g_A^{\nu e}$ & $-0.507 \pm 0.014$ & $-0.507$ & $-0.0$ & $-0.507$ & $-0.0$ \\ 
$(g_L^{eff})^2$ & $0.3001 \pm 0.0014$ & 0.3042 & $-2.9$ & 0.3032 & $-2.2$ \\ 
$(g_R^{eff})^2$ & $0.0308 \pm 0.0011$ & 0.0301 & $+0.6$ & 0.0299 & $+0.8$ \\ 
$Q_W$(Cs) & $-72.18 \pm 0.46$ & $-72.88$ & $+1.5$ & $-72.26$ & $+0.2$ \\ 
$Q_W$(Tl) & $-114.8 \pm 3.6$ & $-116.7$ & $+0.5$ & $-115.7$ & $+0.3$ \\ 
$\sum_{i=d,s,b} |V_{ui}|^2$ & $0.9917 \pm 0.0028$ & 1.0000 & $-3.0$ & 0.9902 
& $+0.5$ \\
\hline
\end{tabular}
\end{center}
\end{table}

We see that we are able to explain the apparent nonunitarity \cite{neutron} 
of the quark mixing matrix and reduce the NuTeV discrepancy \cite{nutev} 
while maintaining excellent agreement with precision data at the $Z$ 
resonance, except for the $b \bar b$ forward-backward asymmetry measured 
at LEP, which is also not explained by the standard model.  In fact, the 
shift of $A_{fb}^{0,b}$ is given in our model by
\begin{eqnarray}
\Delta A_{fb}^{0,b} = {3 \over 4} (A_e \Delta A_b + A_b \Delta A_e) 
= -0.07 \Delta \sin^2 \theta_q - 5.57 \Delta \sin^2 \theta_l.
\end{eqnarray}
Because of the dominant coefficient of the second term, it measures 
essentially the same quantity as $A_l$ and there is no realistic means of 
reconciling the discrepancy of $\sin^2 \theta_{eff}$ at the $Z$ resonance 
using $b \bar b$ versus using leptons in the final state.

The new polarized $e^-e^- \to e^-e^-$ experiment (E158) \cite{e158} at SLAC 
(Stanford Linear Accelerator Center) is designed to measure the left-right 
asymmetry which is proportional to $G_F (1-4\sin^2 \theta_W)$ to an accuracy 
of about 10\%.  Using Eq.~(38) and the standard-model prediction of 
$\sin^2 \theta_W = 0.238$, our expectation is that the above measurement will 
shift by only $-2.2$\% from its standard-model prediction.  The new polarized 
$ep$ elastic scattering experiment (Qweak) \cite{qweak} at TJNAF (Thomas 
Jefferson National Accelerator Facility) is designed to measure $Q_W$ of 
the proton, i.e.
\begin{equation}
Q_W^p = \left[ 1 + \left( {\Delta G_F \over G_F} \right)^{NC}_{lq} \right] 
(1-4s^2)_{SM} - 4 (\Delta \sin^2 \theta_W)_{lq},
\end{equation}
to an accuracy of about 4\%.  We expect a shift of only $+3.0$\%. 
Using Eq.~(38), we see also that the scale of new physics, i.e. $u$ and $w$, 
is at the TeV scale.  Specifically, using the best-fit values of $r$, $y$, 
and $x$, we find $m_{W_2} \simeq m_{Z_2} \simeq 1.2$ TeV, and 
$m_{Z_3} \simeq 0.8$ TeV.

In conclusion, we have proposed a natural gauge model of quark-lepton 
nonuniversality as a foil against the standard model for comparison at 
present and future precision electroweak measurements at both high and 
low energies.  The natural reduction of the effective $G_F$ in lepton-quark 
charged-current interactions versus that in lepton-lepton interactions 
explains the apparent nonunitarity of the quark mixing matrix.  Resulting 
shifts in other electroweak parameters are nontrivial, but a good fit to 
all precision measurements at the $Z$ resonance is still obtained, with a 
small reduction in the observed discrepancy of low-energy neutrino-quark 
scattering data with the theoretical expectation.  New physics at the TeV 
scale is mandatory in the form of one new charged gauge boson and two new 
neutral gauge bosons.

The work of X.L. was supported in part by the China National Natural Science 
Foundation under Grants No.~19835060 and No.~90103017.  The work of E.M. was 
supported in part by the U.~S.~Department of Energy under Grant 
No.~DE-FG03-94ER40837.

\newpage
\bibliographystyle{unsrt}

\end{document}